\documentclass{optica-article}

\journal{opticajournal} 

\articletype{Research Article}

\usepackage{lineno}

\begin{document}

\title{Simple strategy for simulation of large area of axially symmetric metasurfaces}

\author{Augusto Martins,\authormark{1,2*} Achiles F. da Mota,\authormark{3} Chris Stanford ,\authormark{2} Taylor Contreras, \authormark{2} Justo Martin-Albo,\authormark{4} Alexander Kish,\authormark{5} Carlos Escobar, \authormark{5} Adam Para, \authormark{5} and Roxanne Guenette\authormark{1}}

\address{\authormark{1} Department of Physics, University of Manchester, Manchester M13 9PL, United Kingdom\\
\authormark{2} Department of Physics, Harvard University, Cambridge, MA 02138, U.S.A.\\
\authormark{3} Department of Electrical and Computer Engineering, University of Brasília, DF, CEP 70910-900, Brazil\\
\authormark{4} Instituto de F\'isica Corpuscular (IFIC), CSIC \& Universitat de Val\`encia, Calle Catedr\'atico Jos\'e Beltr\'an, 2, Paterna, E-46980, Spain\\
\authormark{5} Fermi National Accelerator Laboratory, Batavia, IL 60510, U.S.A}

\email{\authormark{*}augusto.martinezalves@manchester.ac.uk} 


\begin{abstract*} Metalenses are composed of nanostructures for focusing light and have been widely explored in many exciting applications. However, their expanding dimensions pose simulation challenges. We propose a method to simulate metalenses in a timely manner using vectorial wave and ray tracing models. We sample the metalens’ radial phase gradient and locally approximate it by a linear phase response. Each sampling point is modeled as a binary blazed grating, employing the chosen nanostructure, to build a transfer function set. The metalens transmission or reflection is then obtained by applying the corresponding transfer function to the incoming field on the regions surrounding each sampling point. Fourier optics is used to calculate the scattered fields under arbitrary illumination for the vectorial wave method and a Monte Carlo algorithm is used in the ray tracing formalism. We validated our method against finite difference time domain simulations at 632 nm and we were able to simulate metalenses larger than 3000$\lambda_0$ in diameter on a personal computer.

\end{abstract*}

\section{Introduction}
Metalenses consist of subwavelength nanostructures that locally modify the phase profile of an incoming beam to focus it \cite{ref1,ref2,ref3}. The unprecedented degree of light manipulation, compactness, and compatibility with standard nanofabrication processes make them attractive substitutes for conventional refractive optics systems. These cutting-edge devices have been designed for a wide range of wavelengths (from 50 nm to 3600 nm) and are now developed for industrial applications \cite{ref3}. Diffraction limited focusing \cite{ref4}, wide field of view imaging \cite{ref6,ref7}, achromatic focusing \cite{ref8}, and endoscopic imaging \cite{ref11} are a few examples of the breadth of applications enabled by metalenses. 

Recently, mass-manufacturable metalenses with diameters on the order of a few centimeters have been demonstrated \cite{ref12,ref13,ref14,ref15}. Normally, metalenses are simulated under certain approximations or advanced techniques that reduce the computational burden. Conventional rigorous numerical methods to solve Maxwell’s equations, such as finite difference time domain (FDTD) and finite elements method (FEM), are not suitable because they require enormous computational resources \cite{ref16}. Several strategies to accelerate these methods have been proposed, including hardware acceleration for the FDTD method \cite{ref17}, augmented partial factorization (APF) \cite{ref21}, and low-overhead distribution on a GPU-based simulation \cite{ref22}. These approaches have successfully simulated metalenses with dimensions as large as 600$\lambda_0$ \cite{ref22}. However, the most common approach to simulate even larger area metalenses is the local approximation method \cite{ref7,ref17,ref18,ref19}, where the transmitted field by each nanopost is assumed to be constant and equal to its array response. This approach, however, does not account for the coupling among nanostructures \cite{ref1} and cannot simulate metagratings .

Here, we propose a method to accurately simulate large area metalenses in a timely manner. Our strategy is based on sampling the metalens phase gradient profile and modelling each point as a binary blazed grating using the nanopost or metagrating design of choice. We then build a transfer function library for each blazed grating under plane wave incidence for different incident angles and polarizations. A similar approach has been recently demonstrated to simulate quantum emitters' response near periodic patterned hyperbolic metamaterials \cite{ref24,ref25}. Our model allows us to simulate the metalens response to an arbitrary field distribution under a wave vector and ray tracing models. The model expands the field and rays in terms of the diffraction orders of the metalens, allowing for an unprecedented analysis of the focused field. We  compared our method against rigorous FDTD simulations and managed to reduce the simulation time and memory requirements by at least on order of magnitude whilst keeping the resulst accurate. We used our method to simulate metalenses with diameters larger than 3000$\lambda_0$ and separate the field into the contributions of each diffraction order.

\section{Metalens model rationale}

\begin{figure}[ht!]
\centering\includegraphics[width=5cm]{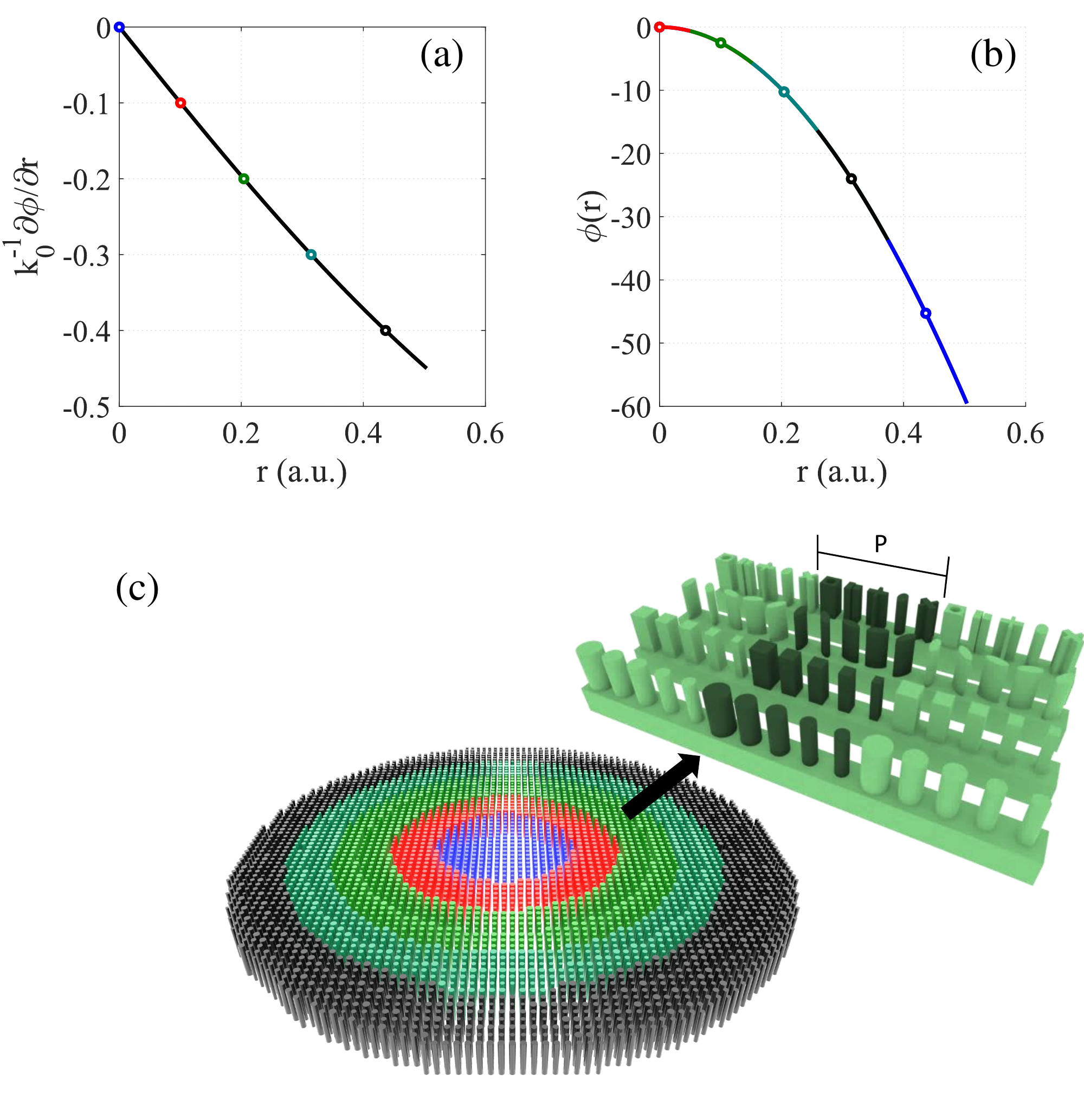}
\caption{(a) and (b) show the sampling of the gradient of an arbitrary radial phase profile and the region they cover on the phase profile, respectively. (c) depicts the equivalent blazed binary grating approximation on the metalens. }
\label{fig:figure1}
\end{figure}

We propose a model based on a transfer-function approach that considers coupling among different posts. The metasurface phase gradient is sampled on $N$ positions, as illustrated in Fig. \ref{fig:figure1}. The $i^{th}$ patch can be modelled as a blazed binary grating with period given by the grating equation as

\begin{equation} 
\label{eq:equation1}
P_i=\dfrac{2\pi}{G(\vec{r}_{i1})}
\end{equation}

\noindent where $G(\vec{r}_{i1})\equiv G_{i1}=\partial\phi(r)/\partial r$ is the radial phase gradient value calculated at the $i^{th}$ sampling point. Any radial phase profile can be sampled using our phase gradient library, as shown in Fig. \ref{fig:figure1}. The blazed binary grating is modeled according to the nanostructure or metagrating design being used such as rectangular, elliptical posts, among others, as shown in Fig. \ref{fig:figure1} (c).  This approximation is solely used to obtain the transmission and/or reflection coefficients of that region and do not account for the phase profile curvature. We address this issue by locally correcting the phase gradient according to the ideal phase profile, as will be discussed later. 

We calculate the blazed binary grating transmission and reflection coefficients under plane wave excitation at different angles of incidence and polarizations using the rigorous coupled-wave analysis (RCWA) method \cite{ref28,ref29}. The transfer function is defined as the diffraction efficiencies or field amplitudes of the transmitted/reflected waves as a function of the incoming in-plane wave-vector.  We solve Maxwell’s equations for a supercell of the blazed binary grating on a region with a cross-section area given by $d \times P_i$, where $d$ is the unit cell size of a single post on the metalens design, as shown in Fig. \ref{fig:figure1} (c). Note that depending on the metasurface nanopost unit cell size, we may have to use a supercell larger than $P_i$ to fit an integer number of posts inside it. With this approach, the coupling among the posts is fully modeled, but the coupling between different regions is neglected. To generate each blazed binary grating, we perform the phase profile sampling as necessary.  After calculating the transfer functions, they are used to obtain the scattering properties of the metalens.   The transfer function calculation needs to be realized once,  and it can be hastened using accurate and faster methods \cite{ref17,ref21,ref22}. To save computation effort, we calculate a single transfer function for each radial sector to later rotate them to the new system of coordinates taking advantage of the metalens rotation symmetry.

\subsection{Vectorial wave model}
\begin{figure}[ht!]
\centering\includegraphics[width=6cm]{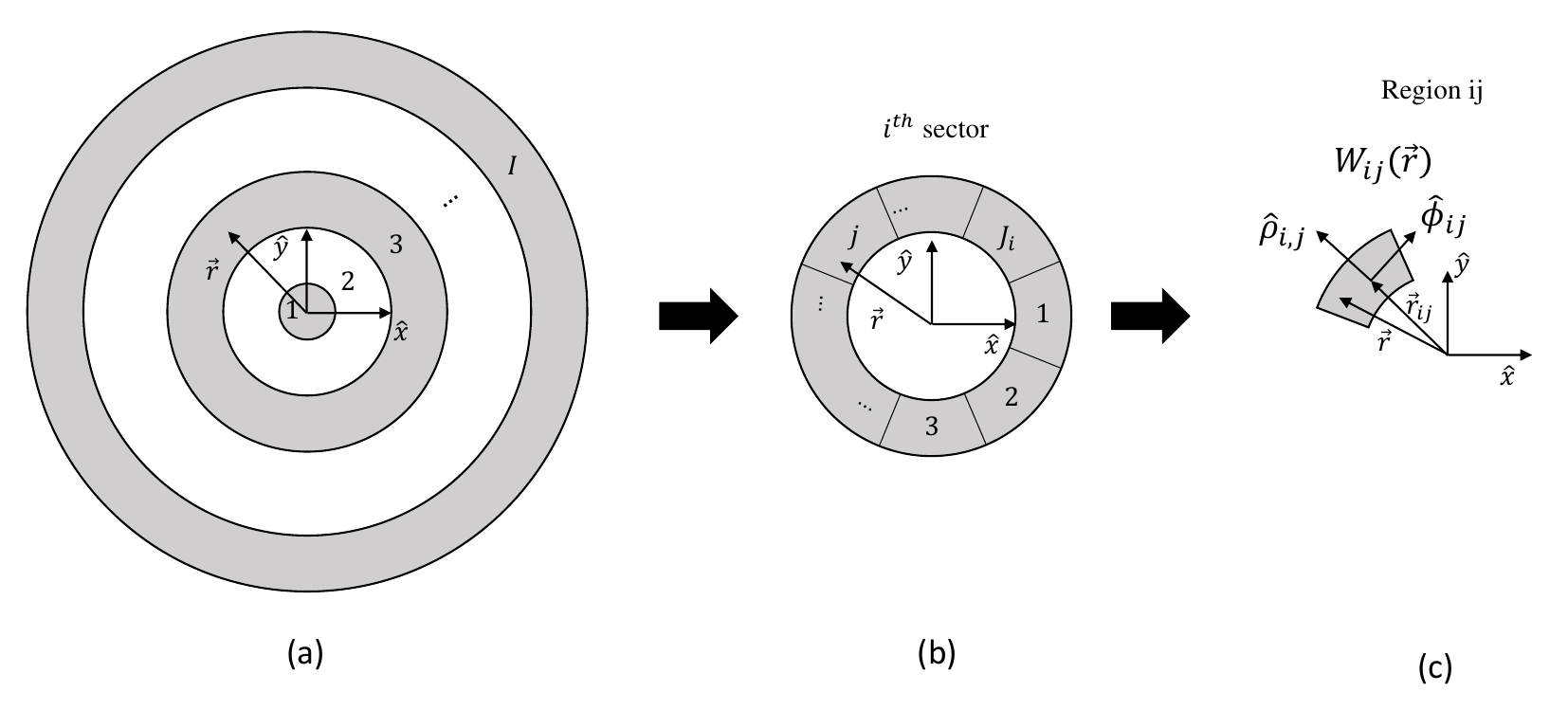}
\caption{(a) show how the different sectors are set up based on the gradient sampling. (b) shows the angular splitting for each sector, forming a given region.   $\vec{r}$ is defined as the position vector centered on the metalens and $\vec{r}_{ij}$ is the vector pointing to the center of the region ij. $\hat{\rho}_{ij}$ and $\hat{\phi}_{ij}$ are the blazed direction versor and the corresponding orthogonal versor of region ij, respectively.  }
 \label{fig:figure2}
\end{figure}

The metalens is modeled as a spatially patched transfer function that modulates an arbitrary field distribution and relies on the angular spectrum formalism for the propagation in free space \cite{ref26}. As explained in the previous section, we sample the metalens phase profile gradient with linear patches as shown in Fig. \ref{fig:figure2}(a). Moreover, we also split the metalens along the azimuthal direction, leveraging its rotation symmetry to use the same transfer function but properly spatially rotated, as shown in Figs. \ref{fig:figure2}(b) and \ref{fig:figure2}(c). That is, we approximate the metalens phase profile by a piece-wise linear phase profile at the points $\vec{r}_{i,j}$  creating $I$ sectors radially, and we split each sector into $J_i$ regions where $i$ indicates a given radial sector, as represented in Fig. \ref{fig:figure2}.  The total transmitted or reflected fields by the metalens are calculated as the sum of the contribution of each sector
\begin{equation}
\label{eq:equation2}
    \vec{E}(\vec{r})=\sum_{i=1}^I\sum_{j=1}^{J_i}\vec{E}_{ij}(\vec{r}){W}_{ij}(\vec{r})
\end{equation}
where $\vec{E}_{i,j}(\vec{r})$ is the electric field distribution calculated on sector $j$ of region $i$ and ${W}_{ij}(\vec{r})$ is a window function, as shown in Fig. \ref{fig:figure2}(c), that limits the region area and is given by Eq. (S2). $\vec{r}$ is defined as a radial vector centered on the metalens (see Fig. \ref{fig:figure2} (a))  .
If the incoming field distribution has a Fourier transform given by $\boldsymbol{\vec{E}_{0}}(\vec{k}_{\parallel})$ (bold symbols represent the Fourier transformed quantities), the transmitted field spatial spectrum on axial region j of sector i (see Fig. \ref{fig:figure2}) is given by 
\begin{equation}
\label{eq:equation4}
    \vec{\boldsymbol{E}}_{ij}(\vec{k}_{\parallel}) = \int d\kappa^{2} \stackrel{\leftrightarrow}{T}_{ij}(\vec{k}_{\parallel},\vec{\kappa}_{\parallel}) \cdot \boldsymbol{\vec{E}_{0}}(\vec{\kappa}_{\parallel})
\end{equation}
 where $\stackrel{\leftrightarrow}{T}_{ij}(\vec{k}_{\parallel},\vec{\kappa}_{\parallel})$ is the transfer function defined as a tensor given by Eq. (S4). Eq. (\ref{eq:equation4}) assumes that the metalens acts as a linear operator on the incoming field. Such ansatz is corroborated by the linear property of Maxwell's equations \cite{ref24,ref25,ref26}. As shown in the SI, after operating the transfer function on the incoming field distribution, the transmitted field spectrum can be calculated as
\begin{equation}
\label{eq:equation9}
    \boldsymbol{\vec{E}_{ij}}(\vec{k}_\parallel) = \sum_g \vec{\boldsymbol{e}}_{ijg}(\vec{k}_{\parallel}-\vec{G}_{ijg})
\end{equation}
where $\vec{\boldsymbol{e}}_{ijg}(\vec{k}_{\parallel})$ is the spectrum of the output field produced by the $g$-th diffraction order and is given by Eq. (S10). The reciprocal vectors are defined according to the versors $\hat{\rho}_{ij}$ and $\hat{\phi}_{ij}$, shown in Fig. \ref{fig:figure2} (c). Finally, to obtain the field in real space, we simply perform an inverse Fourier transform in  $\vec{\boldsymbol{e}}_{ijg}(\vec{k}_{\parallel})$. Applying the inverse Fourier transform and its shifting property in Eq. (\ref{eq:equation9}) we have that 
\begin{equation}
    \label{eq:equation11}
    \vec{E}_{ij}(\vec{r}_\parallel)= \sum_g e^{-j \vec{G}_{ijg}\cdot \vec{r}_\parallel} \int dk^2_\parallel \vec{\boldsymbol{e}}_{ijg}(\vec{k}_{\parallel}) e^{j \vec{k}_{\parallel}\cdot \vec{r}_\parallel}
\end{equation}
As seen in Eq. \ref{eq:equation11}, the total field on the $(i,j)$ region is given by the linear superposition of the diffracted fields modulated by the linear phase of the order. Furthermore, the phase term $e^{-j \vec{G}_{ijg}\cdot \vec{r}_\parallel}$ controls the central position of the diffracted field spectrum and creates a local  linear phase profile distribution, which is a consequence from the linear patching approach. This effect induces aberrations on the wavefront and can be detrimental to the properties of the metalens. To eliminate this problem, we correct this phase term by substituting it to the original phase profile, $\phi(\vec{r}_\parallel)$, as discussed in the SI.
\subsection{Ray tracing}
Although more precise and faster in describing the behavior of large-area metasurfaces, the computational burden of the vectorial approach can be further reduced by applying the proposed method on the ray tracing method. The metalens is now treated as a phase discontinuity on the ray path \cite{ref30} but also accounting for the diffraction efficiency of the metalens. Given a ray with in-plane momentum $\vec{\kappa}_\parallel$ incoming at a point $\vec{r}_\parallel$ on the metalens (see Fig. \ref{fig:figure2}(a)), we can model the diffraction by treating it as a probabilistic event with a Monte-Carlo algorithm. That is, there are three possible outcomes for the ray in our model after it interacts with the metalens: it can either be diffracted back (reflection), forward (transmission), or absorbed if the structure is lossy. The probabilities are taken as the diffraction efficiencies of the patch approximation obtained by solving Maxwell’s equations. For the scattering processes we define $T_G(\vec{k}_\parallel,\vec{P})$ and $R_G(\vec{k}_\parallel,\vec{P})$ as being the probabilities of a ray with incoming in-plane momentum $\vec{k}_\parallel$ and polarization state $\vec{P}$ being diffracted in transmission and reflection, respectively, and $G\in \mathbb{D}$ is a given diffraction order in the set of stored diffraction orders space ($\mathbb{D}$). From energy conservation, we can define the probability of a ray being absorbed as $A$, as shown in the SI.

We define a cumulative probability distribution $f$ as
\begin{equation}
\label{eqRT:equation7}
      f[i]=\begin{cases}
      \sum_{j=1}^i V[j] & 1\leq i \leq M+N+1 \\
      0 & i=0
      \end{cases}
\end{equation}
where, $V[i]\equiv[T_{G_1},T_{G_2},\cdots T_{G_N},R_{G_1},R_{G_2},\cdots R_{G_M},A]$ and we omitted the region indices for clarity. Therefore, we can use $f$ to define intervals where a given outcome might happen. The diffraction order can then be found by generating a random number and analysing in which interval it has fallen. Given the uniformly distributed random variable $\chi\in[0,1]$, we can obtain $i$, and consequently the outcome of the event from the sequence $V$, by solving the following equation
\begin{equation}
\label{eqRT:equation8}     g(\chi)=i, \text{ if } f[i-1]<\chi\leq f[i]
\end{equation}
where $g$ is the resulting diffraction index and index of the sequence $V$, which is used to map a given event. With the scattering event determined, the resulting ray can either be scattered or absorbed. If it is absorbed, then no other ray is produced. On the contrary, if the ray
is reflected or transmitted, then a new ray is generated with momentum given by
\begin{equation}
\label{eqRT:equation9}   \vec{k}=\vec{\kappa}_{\parallel}+\vec{G}_i\pm \hat{n}k_n
\end{equation}
where $\hat{n}$ is the normal vector to the metasurface, $\vec{G}_i$ is the reciprocal vector of the g-th order, and $k_n$ is the resulting ray wave-vector along the normal direction, which can be found from the dispersion equation in the medium. Note that the reciprocal vector radial component is corrected locally according to the local phase gradient instead of using the blazed binary grating momentum.  As shown in the SI, we can also obtain the diffracted polarization by the metalens. Each ray can be labelled according to the corresponding diffraction order it originated from, allowing a better understanding of the metalens behavior, without the need to perform separate simulations, see Figs. (S1) and (S2) in the SI for more detail. Such discrimination is also possible on the vectorial wave method, but it would require storing a field distribution of each order, which is impractical. 
\section{Results}
\subsection{Focusing profile comparison}

\begin{figure}[ht!]
\centering\includegraphics[width=8cm]{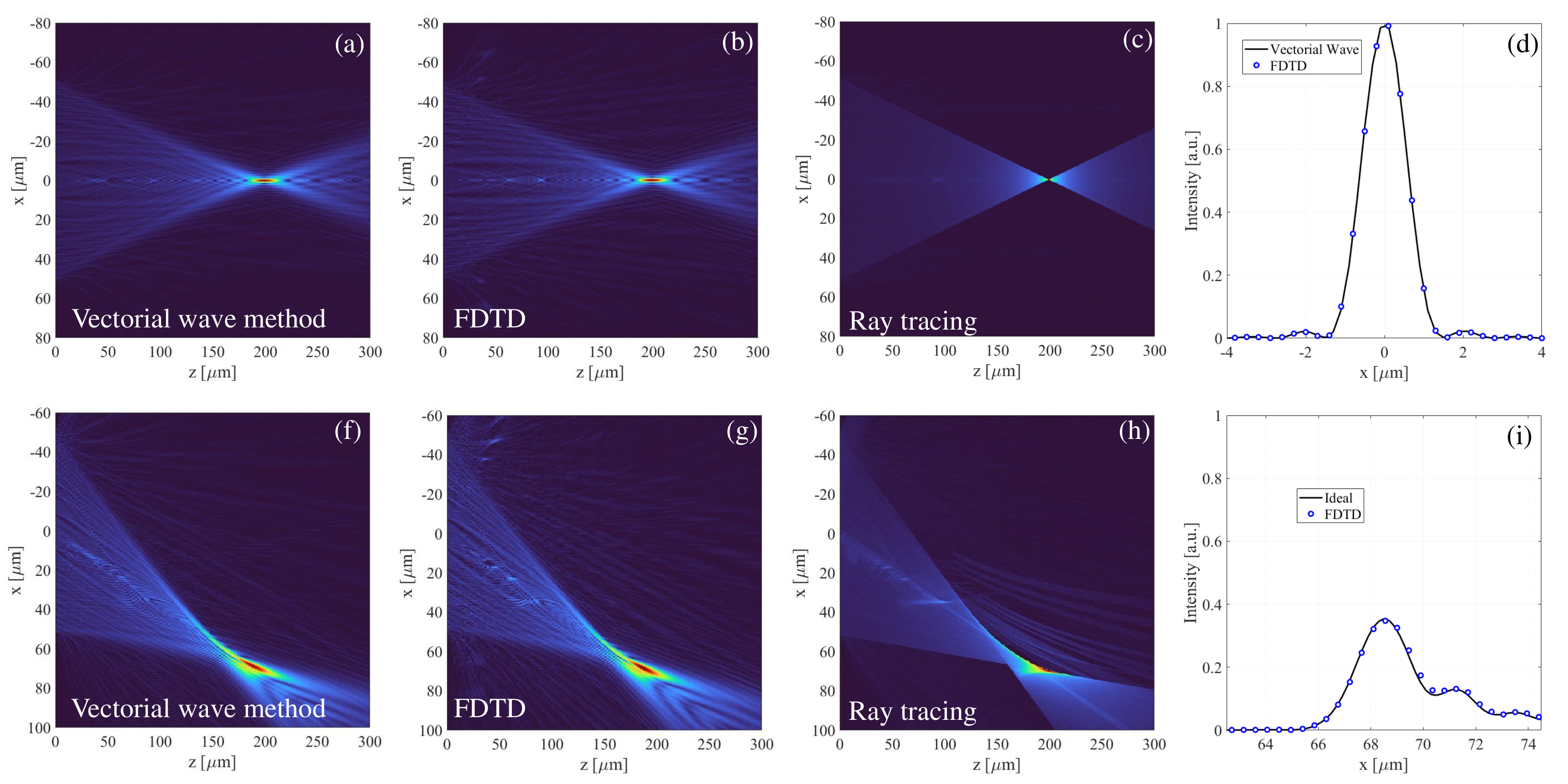}
\caption{Field intensity distribution focused by a metalens with a focal length of 200 $\mu$m and NA=0.25 calculated using different methods. The operating wavelength is 632 nm in all cases, (a)–(d) [(f)–(i)] show the longitudinal field amplitude distributions focused by the metalens at normal [$20^\circ$] incidence calculated using the vectorial wave linear patch, FDTD, ray tracing with linear patching.  (e) and (j) show the transversal cuts of the field distributions on the focal plane at normal and $20^\circ$ incidence, respectively.}
\label{fig:figure5}
\end{figure}

 We  compare the focusing profile of a metalens operating at 632 nm against FDTD simulations. The ray tracing map distribution is obtained by calculating the ray density crossing the plane $y=0$ and the field distributions were normalized with respect to their peak values on the focal plane to highlight the field distribution. The simulation parameters are discussed in section \ref{section:computational_resources}.  We use  1200 nm tall glass-based nanoposts in our design \cite{Taylor,ref15}.  The metalens focal length is 200 $\mu$m with a diameter of 100 $\mu$m (NA = 0.25) and it encodes a hyperbolic phase profile. 
The vectorial wave based model qualitatively reproduces the FDTD field distribution with good fidelity, even accounting for the appearance of higher-order focal spots as shown in Figs. \ref{fig:figure5}(a) and  \ref{fig:figure5}(b), respectively, at normal incidence. The model can also reproduce the field distribution at 20$^o$ of incidence as shown by Figs.  \ref{fig:figure5}(e) and  \ref{fig:figure5}(f), respectively. The ray tracing model also simulates the focusing field distribution and the high order focal spots as shown in Figs.  \ref{fig:figure5}(c) and  \ref{fig:figure5}(g). However, it accounts only for the power flow of the diffracted rays and idealizes the intensity distribution as it does not model interference among the rays. The interference leads to the well-known diffraction limit in optics and can be visualized by the field intensity distribution on the focal plane, as shown in Figs.  \ref{fig:figure5}(d) and \ref{fig:figure5}(h)  at normal and oblique ( $20^\circ$ )incidence, respectively.  All plots were normalized with to their own power flux and peak intensity of the ideal Airy disk distributrion. The field distributions at the focal plane obtained by our method presents good agreement with the corresponding distributions FDTD at normal and oblique incidences. The transversal cuts of the point spread function at oblique incidence are highly distorted due to off-axis aberration – mainly coma, as shown in Fig.\ref{fig:figure5}(j). We also calculated the focusing efficiency on an circle with diameter ten times larger then the PSF full width at half maximum (FWHM). At normal incidence, we obtained 73.5\%,  72.3\% and 71.5\%   with the FDTD, vectorial wave and ray tracing methods respectively. 

\begin{figure}[ht!]
\centering\includegraphics[width=6cm]{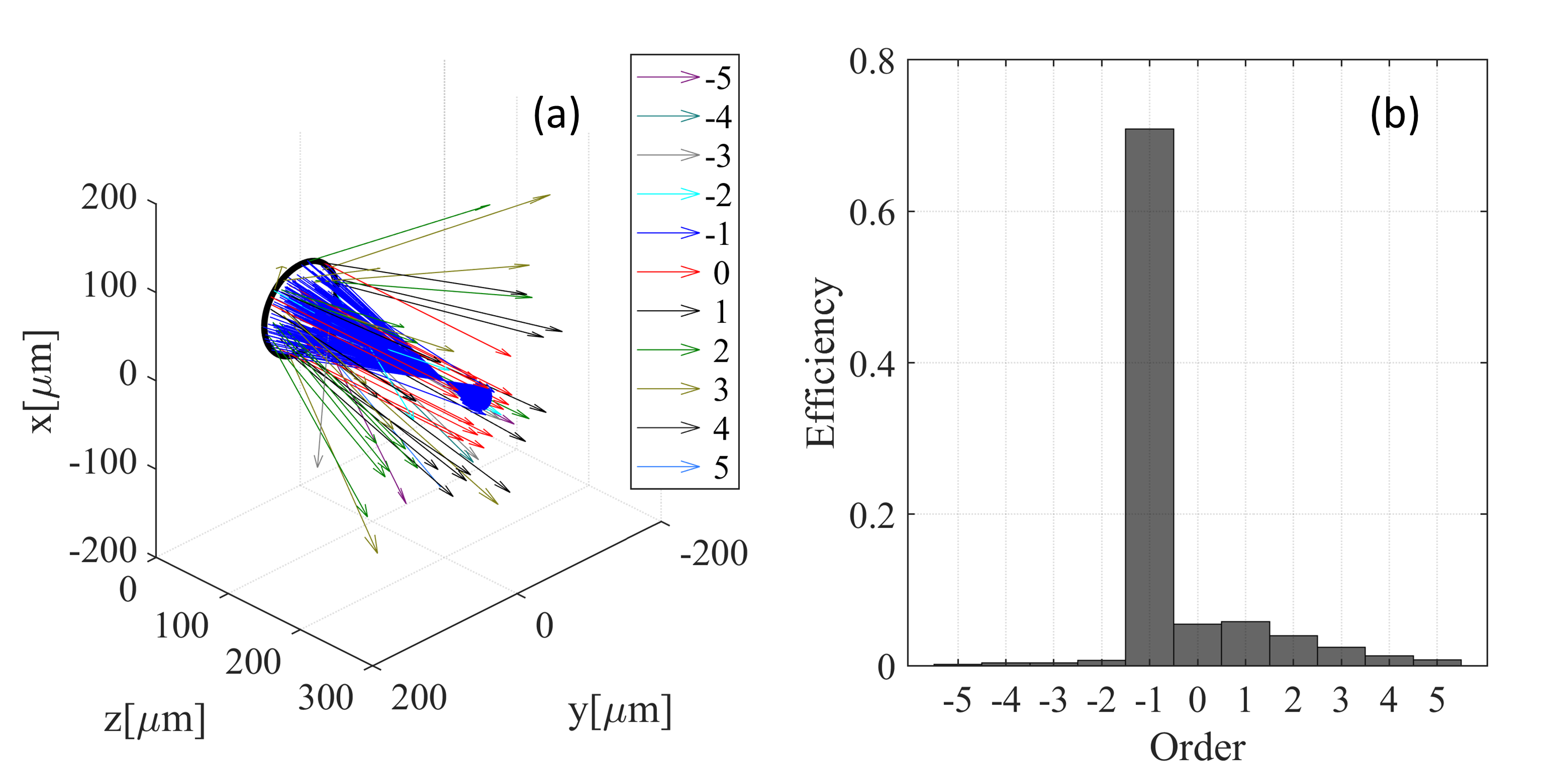}
\caption{Ray tracing of a hyperbolic metalens with $f=200\mu m$ and NA=0.25. (a) and (c) show the resulting ray tracing for the metalens illuminated at normal and oblique (20$^o$) incidence, respectively. We only show 250 rays on this plot and limited the diffraction order to $\pm 5$.  (b) and (d) show the diffraction orders normalized histograms of the transmitted rays for normal and oblique incidences, respectively.  }
\label{fig:raytracing}
\end{figure}

Finally, the ray tracing model allows us to easily tag each ray according to the diffraction order it originated. Fig. (\ref{fig:raytracing})(a) shows the ray tracing of the simulated metalens at normal incidence. To avoid overcrowding, the plot is limited to 250 rays in total and discriminates the diffraction orders according to the ray color, which are limited to the $[-5,5]$ range. The negative orders gives rise to additional short real focal spots, whereas the positive ones to virtual focal spots. We can calculate the diffraction efficiencies of each order by tallying the number of rays directed into them. Fig.  \ref{fig:raytracing}(b) shows the resulting normalized diffraction orders histograms produced at normal and oblique incidence, respectively. We also applied our model to simulate the performance of a doublet metalens based on \cite{doublet}, as shown in Fig. S1 of the SI.

\subsection{Diffraction efficiency calculation}
After qualitatively showing that our models can calculate the focusing field profile of a metalens, we assess the efficiency prediction of our model against numerical simulations performed using FDTD. We used a  glass-based nanopost design described in \cite{Taylor} where metagratings are used to increase the metalens numerical aperture. Here, we scan a beam across a metalens and calculate the diffraction efficiency of the first few orders as a function of its position from the metalens center. The metalens focal length is 7 mm with a diameter of 15 mm. We use an unpolarized Gaussian beam with 100 $\mu$m (158 $\lambda$) waist. The simulation region for each scan point is  200 $\mu$m $\times$ 200 $\mu$m. 

\begin{figure}[ht!]
\centering\includegraphics[width=12cm]{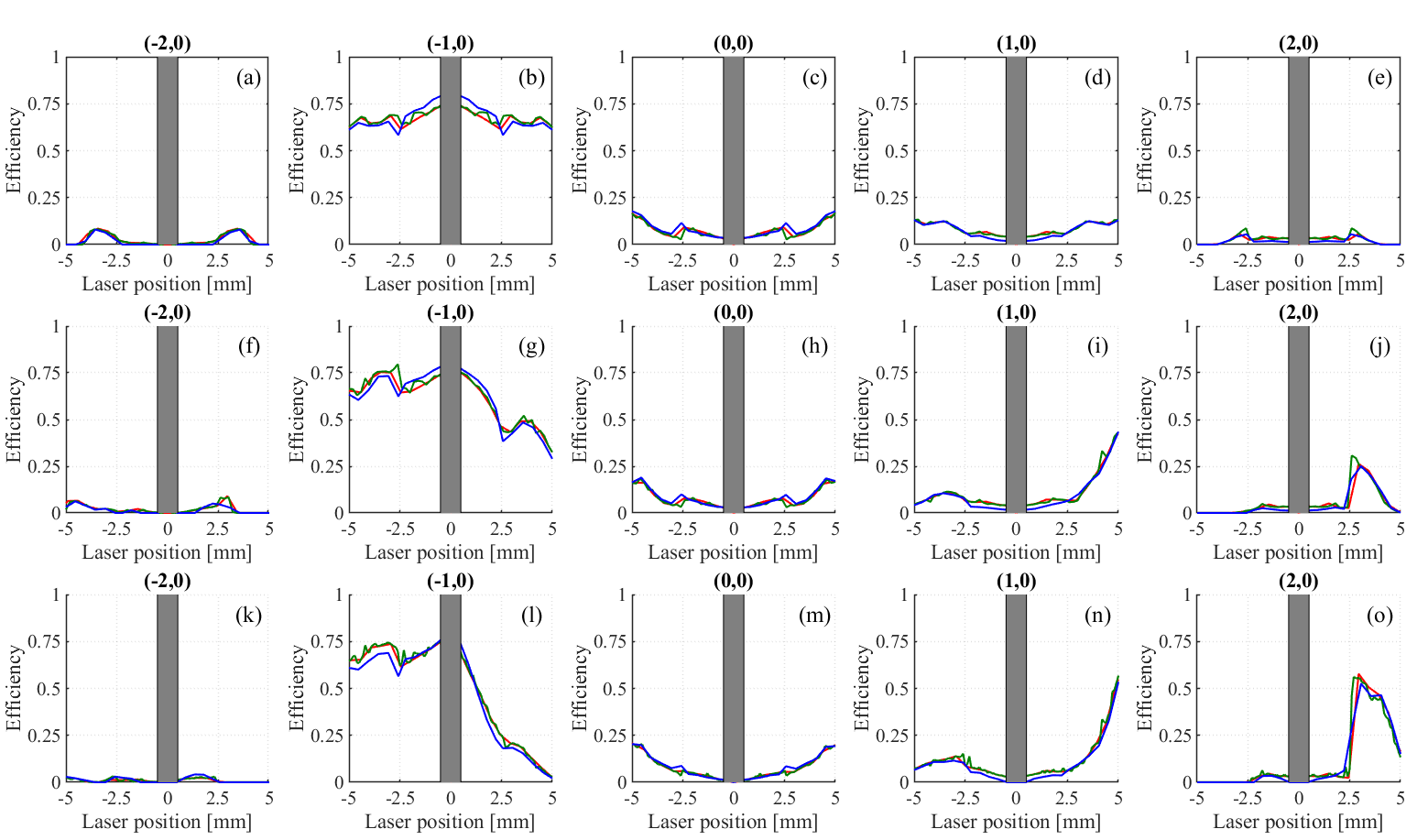}
\caption{Diffraction efficiencies of different orders scattered by the metalens when scanned radially by an unpolarized Gaussian beam with 100 $\mu$m of waist and different angles of incidence along the scanning line. From left to right, the columns show the diffraction efficiency of the -2,-1,0,1, and 2 orders, respectively.  The first, second and third rows show, respectively, the results at normal incidence, $10^\circ$ and $20^\circ$ of incidence. The green, red, and blue lines show the ray tracing model, our vectorial wave model, and the FDTD model results. The operating wavelength is 632 nm. The metalens focal length is 7 mm with a diameter of 15 mm.}
\label{fig:figure4}
\end{figure}

The simulation results at normal, $10^\circ$ and $20^\circ$ of incidence are shown in  (Figs. \ref{fig:figure4}(a)-\ref{fig:figure4}(c)),  Figs. \ref{fig:figure4}(d)-\ref{fig:figure4}(f) and Figs. \ref{fig:figure4}(g)-\ref{fig:figure4}(i), respectively. The calculated diffraction efficiencies of orders m=-2,-1,0 and m=2 and n=0 correspond to each column of Fig. \ref{fig:figure4}, where the metalens focusing comes from the m=-1 order. All methods agree for all angles of incidence and diffraction orders. In particular, the efficiency of the $m=-1$ order is the highest at normal incidence, and it is almost constant throughout the whole scan, remaining higher than 60$\%$, which is an good indication that the proposed design focuses light efficiently.  These results highlight how accurately our model can simulate the metalens even at oblique incidence when coupling among adjacent posts takes over. These results have also been compared against experimental data with a good match, see \cite{Taylor}.   

\subsection{Computational resources} \label{section:computational_resources}

\begin{figure}[ht!]
\centering\includegraphics[width=7cm]{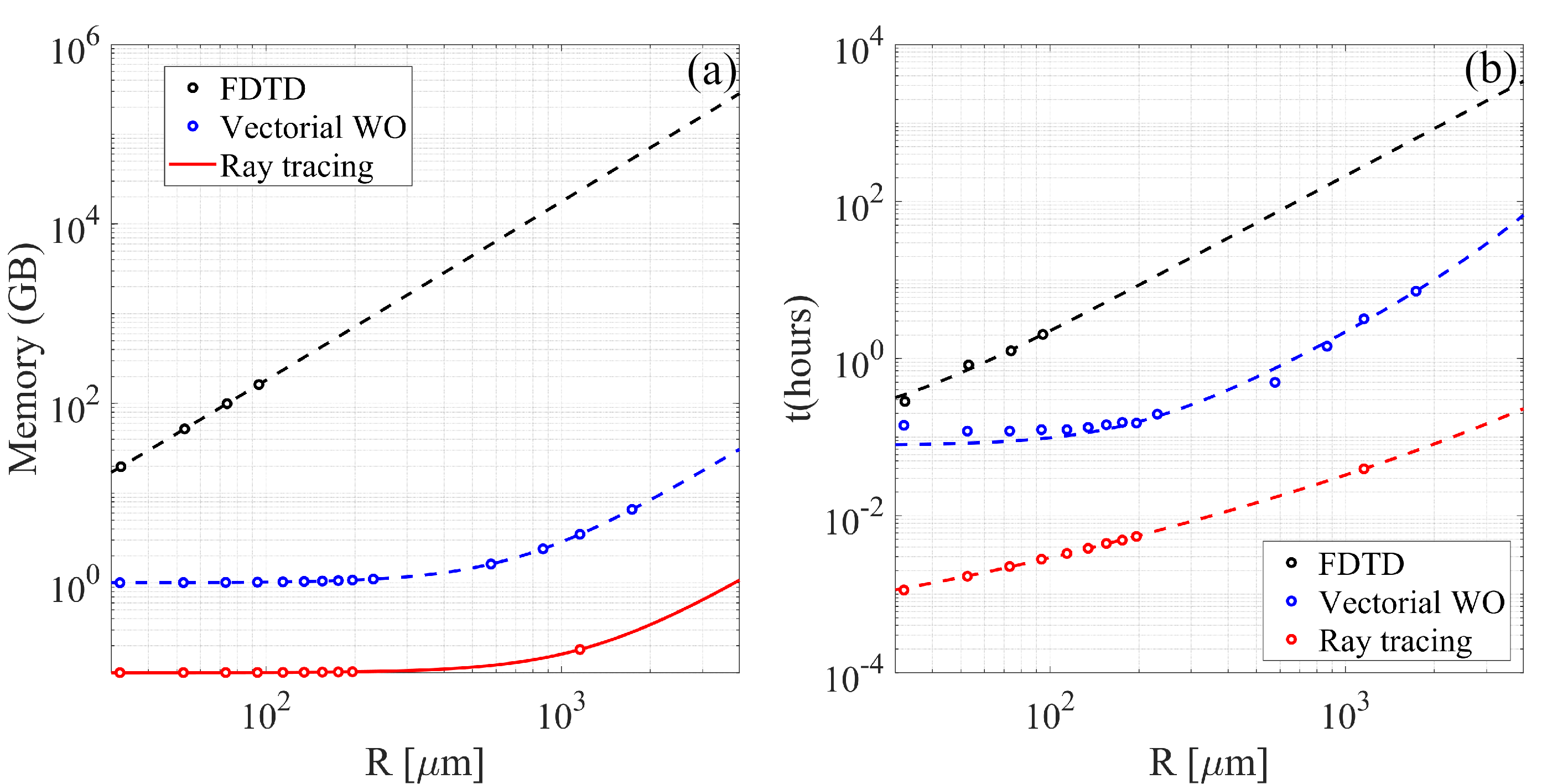}
\caption{ (a) and (b) show, respectively, the memory and time required to simulate glass-based metalenses with different radius ($R$) (dots) and the estimated values using polynomial interpolation from the computed data (dashed lines). The FDTD simulations are used only to calculate the transmitted near field and are performed using Lumerical on the FAS cluster with 128 cpus distributed over 4 nodes using parallelization. The other models were run in serial and have room for improvement. All simulations are performed on a normal incident plane wave operating at 632 nm.}
\label{fig:figure3}
\end{figure}

In this section, we compare each method's simulation time and memory requirements against FDTD simulations. Note that this estimation does not account for the computational resources used in calculating the transfer functions. We simulate fused silica glass-based metalenses, operating at 632 nm \cite{Taylor}, with NA = 0.5 and different focal lengths (f). The FDTD simulations are performed on the commercial software Lumerical inc. using a uniform mesh with 20 nm sampling in all directions. The simulation volume is set to 2R $\times$ 2R $\times$ 2 $\mu m$, where R is the metalens radius $(R = f \tan(\text{asin}(NA))$. Moreover, the FDTD simulations are performed only to calculate the transmitted near field by the metalens, and the free space propagations can be obtained using the angular spectrum formalism \cite{ref26}. The FDTD simulations are carried out on the FAS cluster with 128 cpus distributed over 4 nodes. The proposed approaches are executed in serial on a personal laptop and have room for improvement if a parallelization scheme is used. One could easily distribute the simulation of each patch to different CPUs to hasten the simulation. The field matrices on the wave optics model are calculated using 200 nm of sampling. Finally, the ray tracing model is calculated using a ray density of 5000 rays/$\mu m^2$ hitting the metalens. Figs. \ref{fig:figure3}(a) and \ref{fig:figure3}(b) show the memory required and the time consumed for each method as a function of the metalens radius (increasing focal lengths). Note that even though the models developed here run in serial, the time required for the simulation is considerably lower compared to the FDTD simulation. Furthermore, the proposed approach requires at least one order of magnitude less memory. 

\section{Large area metalens focusing profile simulations}

\begin{figure}[ht!]
\centering\includegraphics[width=12cm]{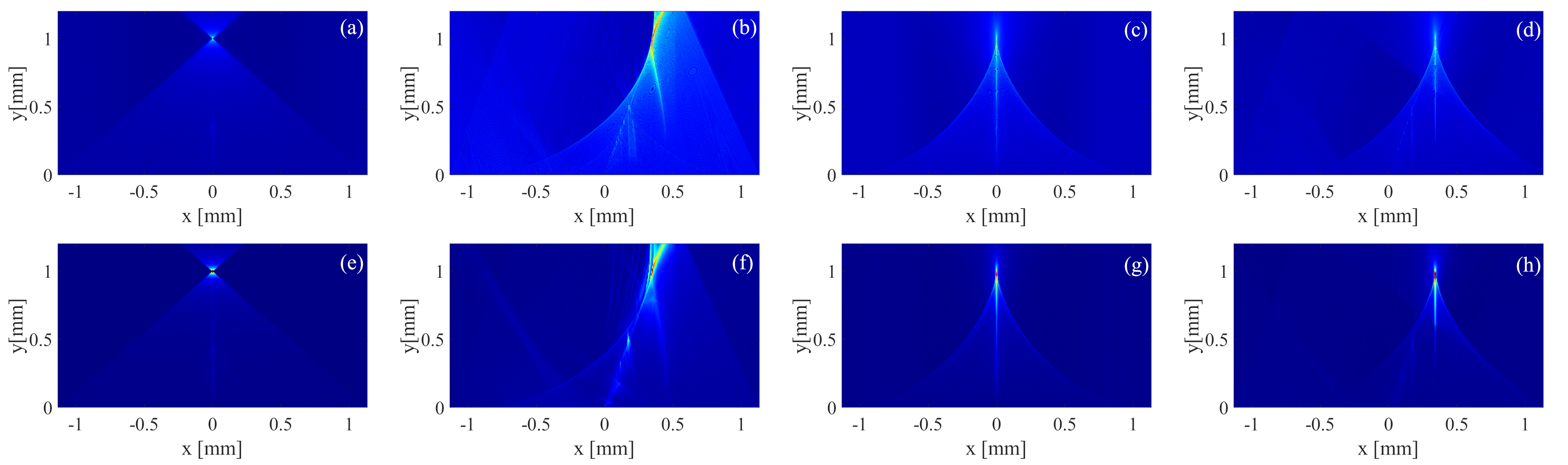}
\caption{(a) - (d) and (e) - (h) show the longitudinal field intensity distribution focused by the metalenses that were calculated using the wave optics and ray tracing models, respectively. (a) and (b) [(e) and (f)] show the focusing profile of a hyperbolic metalens at normal and $20^\circ$ of incidence, respectively. (c) and (d) [(g) and (h)] show the focusing profile of a hyperbolic metalens at normal and $20^\circ$ of incidence, respectively. The metalenses focal length are 1 mm with a diameter of 2.26 mm (NA = 0.75) and the operating wavelength is 632 nm.}
\label{fig:figure6}
\end{figure}

Our approach's low computational time and minimal memory usage enable the simulation of large-area metasurfaces. Here, we simulate the focusing profile of hyperbolic and quadratic metalenses \cite{ref7,ref3} operating at 632 nm, utilizing the same post-design described in the preceding sections. The metalenses focal length are 1 mm with a diameter of 2.26 mm (NA = 0.75), which amounts to  $\sim 3755\lambda_0$. The metalenses are excited by plane waves at normal and $20^\circ$ of incidence. Figs. \ref{fig:figure6}(a)–\ref{fig:figure6}(d) and \ref{fig:figure6}(e)–\ref{fig:figure6}(h) show the longitudinal field intensity distribution focused by the metalenses calculated using the wave optics and ray tracing models, respectively. A metalens with this size would take over 100 hours and require over a petabyte of memory to simulate on a FDTD  cluster with over 400 cpus running in parallel, as shown in Fig. \ref{fig:figure3}(b), only to obtain the near field. Our model requires approximately two orders of magnitude less time and memory to simulate the same metalens.

Both models display similar focused field distributions, but the ray tracing model only accounts for the power flow without any interference effects. At normal incidence, the mirror-symmetric focusing profile of the hyperbolic profile is obtained in both models, and both feature a weak second-order focusing around z=500 $\mu m$, as shown in Figs. \ref{fig:figure6}(a) and \ref{fig:figure6}(e). When illuminated at 20$^o$ of incidence, both models show a strong aberrated focal profile due to coma, and a slightly stronger second-order focusing as shown in Figs. \ref{fig:figure6}(b) and \ref{fig:figure6}(f). The focusing field distributions of the quadratic profile at normal incidence, obtained with the vectorial wave optics and ray tracing methods, are shown in Figs. \ref{fig:figure6}(c) and \ref{fig:figure6}(g), respectively, and do not have mirror symmetry around the focal plane, which is a manifestation of spherical aberration \cite{ref7} . The quadratic field distribution remains almost the same at oblique incidence, as shown in Figs.  \ref{fig:figure6}(d) and  \ref{fig:figure6}(h), due to its wider field of view \cite{ref31}.

\section{Conclusion}

We propose a strategy to simulate large area metalenses by patching it into smaller parts that can be simulated faster using Maxwell's equations. The patching process is similar to the phase sampling used to design a metasurface profile. However, we sample the phase gradient instead, and a corresponding blazed binary grating models each sampling point. Maxwell’s equations are then rigorously solved for each linear piece using a single supercell of the blazed binary grating to obtain a transfer function that is a function of the angle of incidence and polarization. The same transfer function can be reused to model different metalenses or metasurfaces with radial phase profiles for a given metasurface post design. Thus, after the transfer function is obtained, the metalens transmitted or reflected fields can be simulated either using a ray tracing approach or a vectorial wave model, depending on the desired application. We found very good agreement between our model, FDTD simulations, and experiment. This approach reduces the time and memory requirements by orders of magnitude, allowing the simulation of metalenses with diameters larger than 3755$\lambda_0$ on a personal computer.

\begin{backmatter}
\bmsection{Funding}
Content in the funding section will be generated entirely from details submitted to Prism. Authors may add placeholder text in the manuscript to assess length, but any text added to this section in the manuscript will be replaced during production and will display official funder names along with any grant numbers provided. If additional details about a funder are required, they may be added to the Acknowledgments, even if this duplicates information in the funding section. See the example below in Acknowledgements. For preprint submissions, please include funder names and grant numbers in the manuscript.

\bmsection{Acknowledgments}
Most computations in this paper were run on the FASRC cluster supported by the FAS Division of Science Research Computing Group at Harvard University. This document was prepared by using in part the resources of the Fermi National Accelerator Laboratory (Fermilab) and the Noble Liquid Test Facility (NLTF), a U.S. Department of Energy, Office of Science, Office of High Energy Physics HEP User Facility. Fermilab is managed by Fermi Research Alliance, LLC (FRA), acting under Contract No. DE-AC02-07CH11359. 

\bmsection{Disclosures}
The authors declare no conflicts of interest.

\bmsection{Data availability} Data underlying the results presented in this paper are not publicly available at this time but may be obtained from the authors upon reasonable request.

\bmsection{Supplemental document}
See Supplement 1 for supporting content. 

\end{backmatter}

\bibliography{sample}






\end{document}